\documentclass[11pt,numreferences]{kluwer}
\usepackage{klucite}
\begin{document}
\begin{article}

\begin{opening}

\title{A review of the theory of incompressible MHD turbulence }
\author{Benjamin D. G. Chandran \email{benjamin-chandran@uiowa.edu}
\thanks{The author acknowledges the support of NSF grant AST-0098086
and DOE grants DE-FG02-01ER54658 and DE-FC02-01ER54651 at the
University of Iowa.}}  \institute{Department of Physics \& Astronomy, University of Iowa}

\begin{abstract}
This brief review provides an introduction to key ideas in the
theory of incompressible magnetohydrodynamic (MHD) turbulence.
\end{abstract}

\end{opening}

\section{Introduction}
\label{sec:intro} 

Magnetohydrodynamic (MHD) turbulence has been studied in a number of
different parameter regimes, with important contributions from many
authors,
including~\cite{iro63,kra65,pou76,dob80,mon81,gra83,hig84,she83,tin86,zan93,sri94,oug94,gol95,pol95,mon95,ng96,ng97,gol97,gho97,sto98,kin98,bal99,mac99,vaz00,gal00,mul00,cho00,bha01,mar01,lit01,mil01,gal02,oss02,sch02,bol02,cho02,cho03,win03,ves03,pas03,lit03}.
This review focuses on incompressible turbulence in which the
fluctuating velocities are comparable to or less than the~Alfv\'en
speed [defined in equation~(\ref{eq:alfven})].  It is assumed that
there is a mean magnetic field~${\bf B}_0$ that is at least as strong
as the turbulent magnetic-field fluctuations.
Sections~\ref{sec:elsasser} and~\ref{sec:waves} review the MHD
equations and MHD waves, and  section~\ref{sec:wavewave} describes
wave-wave interactions.  Section~\ref{sec:weak}
reviews results on weak MHD turbulence, and section~\ref{sec:strong}
treats strong MHD turbulence.  Section~\ref{sec:dec} describes the
role of dynamic alignment in the decay of MHD turbulence.

\section{Equations of incompressible MHD}
\label{sec:elsasser} 

The governing equations of incompressible MHD are
\begin{equation} 
\rho\left( \frac{\partial {\bf v}}{\partial t} + {\bf v} \cdot \nabla {\bf v} \right)
 = - \nabla \left( p + \frac{B^2}{8\pi} \right) + \frac{1}{4\pi} {\bf B} \cdot 
\nabla {\bf B}  + \rho \nu \nabla^2 {\bf v} ,
\label{eq:mom1} 
\end{equation} 
\begin{equation} 
\frac{\partial {\bf B} }{\partial t} = \nabla \times ({\bf v} \times {\bf B})
+ \eta \nabla^2 {\bf B},
\label{eq:ind} 
\end{equation} 
\begin{equation} 
\nabla \cdot {\bf B} = 0 ,
\label{eq:divb} 
\end{equation} 
\begin{equation} 
\nabla \cdot {\bf v} =0,
\label{eq:divu} 
\end{equation} 
and
\begin{equation} 
\rho = \mbox{ constant},
\label{eq:rho} 
\end{equation} 
where ${\bf B}$ is the magnetic field, ${\bf v}$ is the velocity, $p$
is the thermal pressure, $\rho$ is the density, and~$\nu$ is the
viscosity. In
the limit~$\eta\rightarrow 0$, equation~(\ref{eq:ind}) implies that
field lines are like threads that are frozen to the plasma and
advected at the plasma velocity~${\bf v}$.

Upon defining 
\begin{equation}
{\bf B} = B_0 \hat{\bf z} + {\bf \delta B} ,
\label{eq:mf}
\end{equation} 
where~$B_0 \hat{\bf z}$ is
the uniform mean field,
\begin{equation}
{\bf b} = \frac{{\bf B}}{\sqrt{4\pi \rho} }= 
 v_{\rm A} \hat{{\bf z} } + \delta {\bf b} ,
\label{eq:alfven} 
\end{equation} 
where $ v_{\rm A} = B_0/\sqrt{4\pi \rho}$ is the
Alfv\'en speed, the Elsasser variables
\begin{equation}
{\bf a} ^\pm = {\bf v} \pm \delta {\bf b},
\label{eq:els}
\end{equation}
and the total pressure
\begin{equation}
\Pi = \displaystyle\frac{p}{\rho} +\frac{b^2}{2},
\label{eq:tp} 
\end{equation} 
and upon neglecting the dissipation terms ($\nu=\eta=0$),
equations~(\ref{eq:mom1}) and (\ref{eq:ind}) can be rewritten
as
\begin{equation} 
\frac{\partial {\bf a}^+}{\partial t} - v_{\rm A} \frac{\partial {\bf a} ^+}{\partial z}
= -\nabla \Pi - {\bf a}^- \cdot \nabla {\bf a}^+,
\label{eq:ap} 
\end{equation}
and
\begin{equation} 
\frac{\partial {\bf a}^-}{\partial t}  + v_{\rm A} \frac{\partial {\bf a} ^-}{\partial z}
= -\nabla \Pi - {\bf a}^+ \cdot \nabla {\bf a}^-.
\label{eq:am} 
\end{equation} 

In the absence of dissipation, equations~(\ref{eq:mom1}) through
(\ref{eq:rho}) possess three quadratic invariants: the energy 
$ (1/2) \int (v^2 + b^2)\, d^3 x$, the cross helicity $(1/2)\int
{\bf v} \cdot \delta {\bf b} \,d^3 x= (1/8) \int ({\bf a}^+ \cdot {\bf
a}^+ - {\bf a}^- \cdot{\bf a}^-) d^3 x$, and the magnetic helicity
$\int {\bf A} \cdot {\bf B} \,d^3 x$, where ${\bf A}$ is the magnetic
vector potential. The magnetic helicity measures linkages of magnetic
field lines and is associated with an inverse
cascade~\cite{pou76}. The cross helicity describes the excess of one
fluctuation type (${\bf a} ^+$ or ${\bf a} ^-$) over the other, and is
associated with the phenomenon of dynamic alignment
(section~\ref{sec:dec}).

\section{Waves and wave packets in incompressible MHD}
\label{sec:waves}

If ${\bf v} = \delta {\bf b} $ (i.e., ${\bf a}^- = 0$), the solution
to equation~(\ref{eq:ap}) for $\nabla \times {\bf a} ^+$ is a wave
that travels at speed $v_{\rm A}$ in the $-z$ direction. Similarly, if
${\bf v} = - \delta {\bf b}$ (i.e., ${\bf a}^+ = 0$), the solution to
equation~(\ref{eq:am}) for $\nabla \times {\bf a}^-$ is a wave that
travels at speed $v_{\rm A}$ in the $+z$ direction.  These wave
solutions are valid for arbitrarily large wave amplitude
(when either ${\bf a}^+=0$ or ${\bf a}^- =0$). 

\begin{figure}[h]
\vspace{2in}
\includegraphics{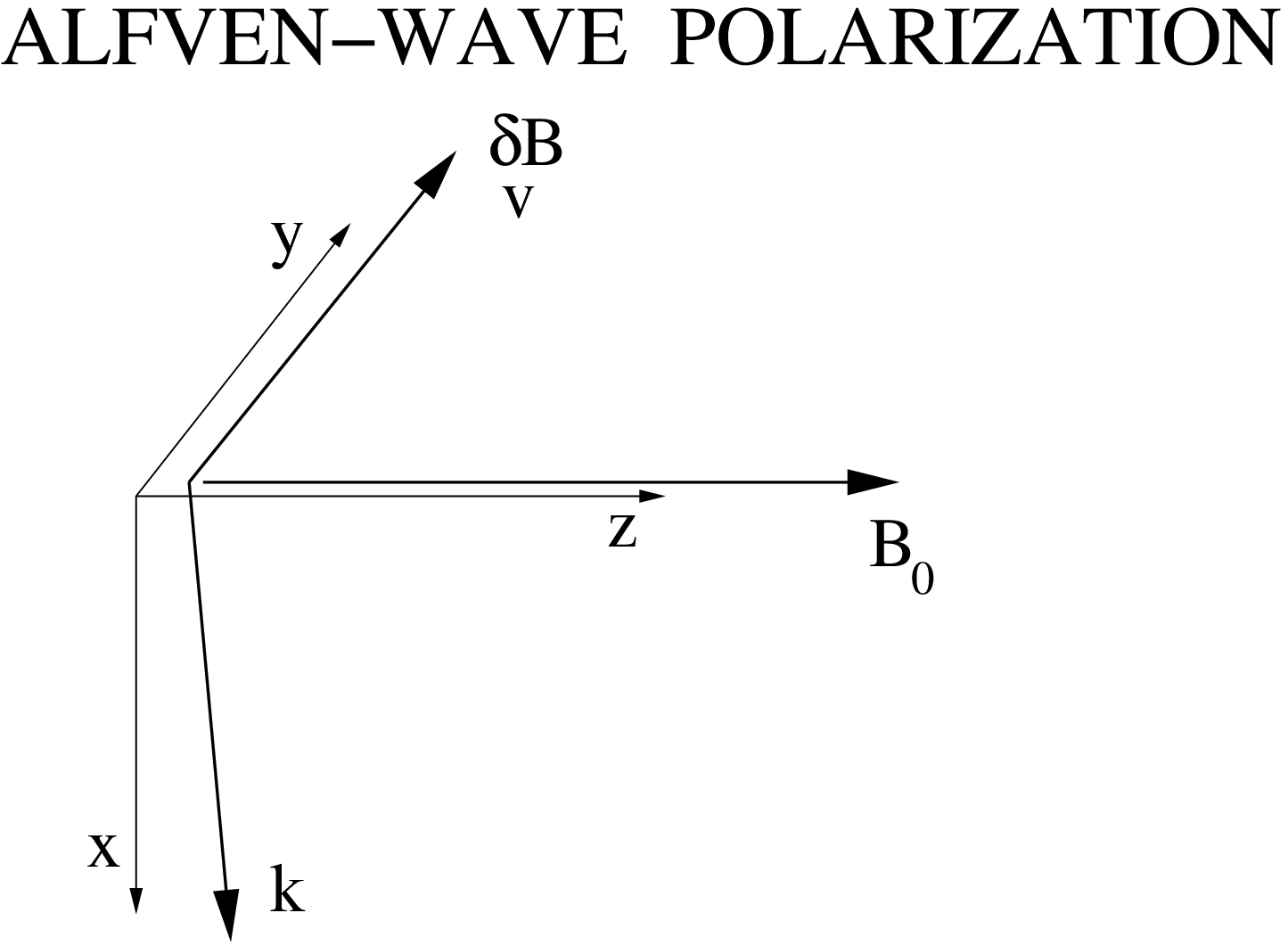}
\includegraphics{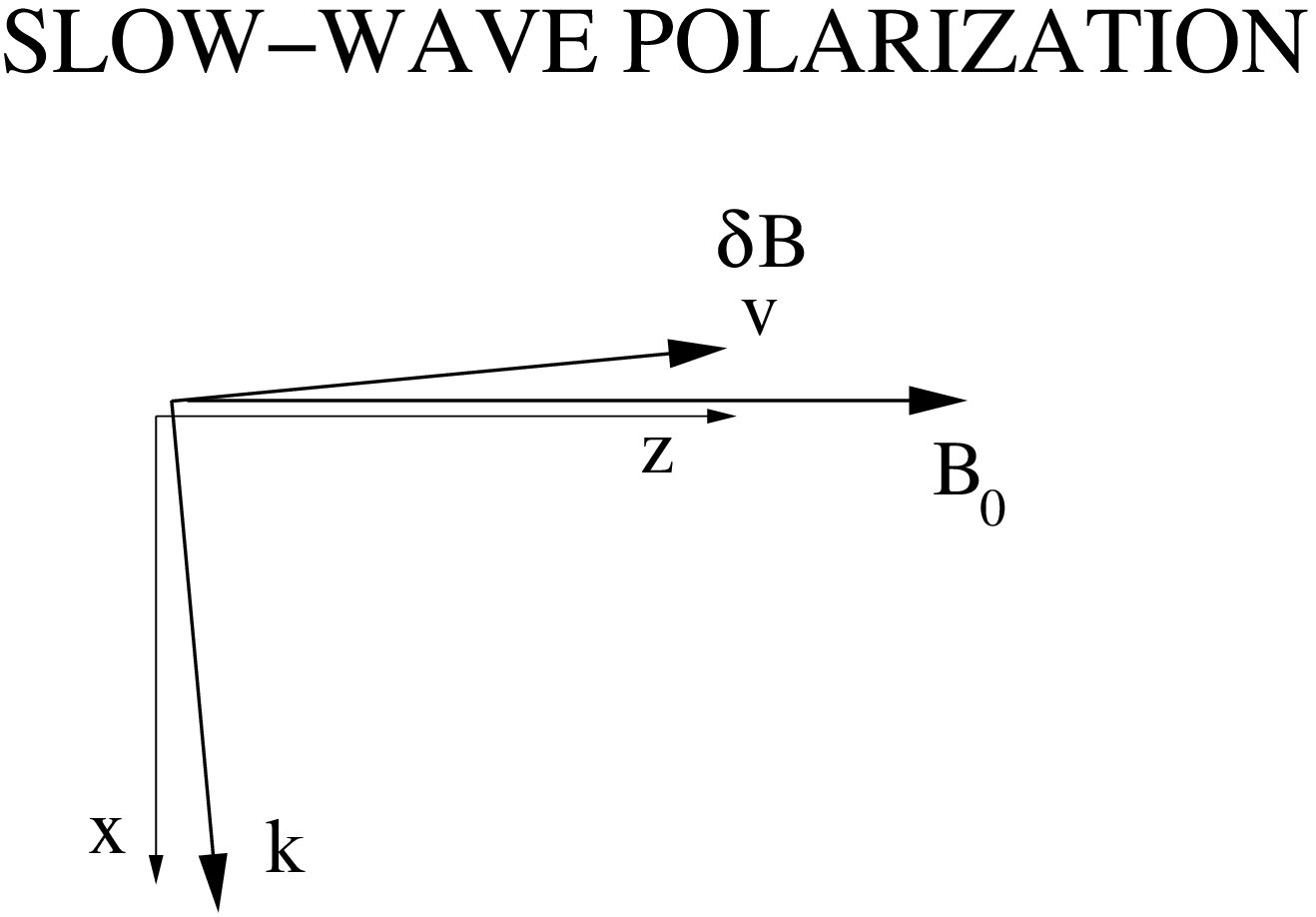}
\caption{For an Alfv\'en wave, $\delta {\bf v}$ and $\delta {\bf B}$ are
$\perp$ to both~${\bf B}_0$ and the wave vector~${\bf k}$. For a slow wave,
$\delta {\bf v}$ and $\delta {\bf B}$ are $\perp$ to ${\bf k}$,
but in the plane of ${\bf k}$ and ${\bf B}_0$. If ${\bf k}$ is nearly
$\perp$ to~${\bf B}_0$, then $\delta {\bf v}$ and $\delta {\bf b}  $
are nearly along ${\bf B}_0$ for a slow wave.}
\label{fig:polarizations}
\end{figure}

\begin{figure}[h]
\vspace{2.5in}
\includegraphics{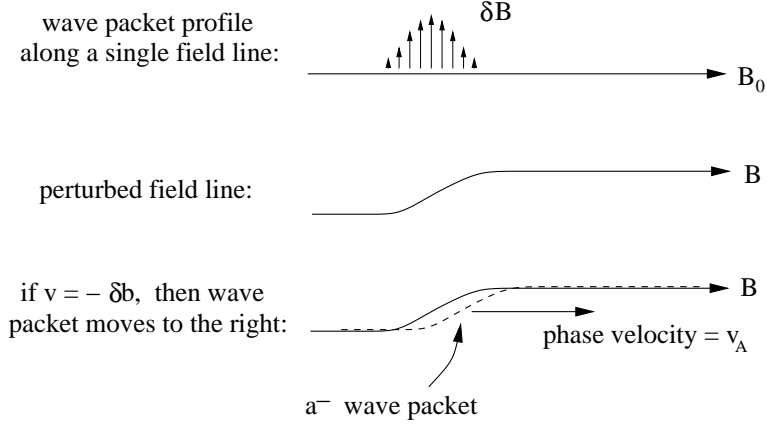}
\caption{The profile of a 1D Alfv\'en wave packet.}
\label{fig:wavepacket1}
\end{figure}

\begin{figure}[h]
\vspace{2in}
\includegraphics{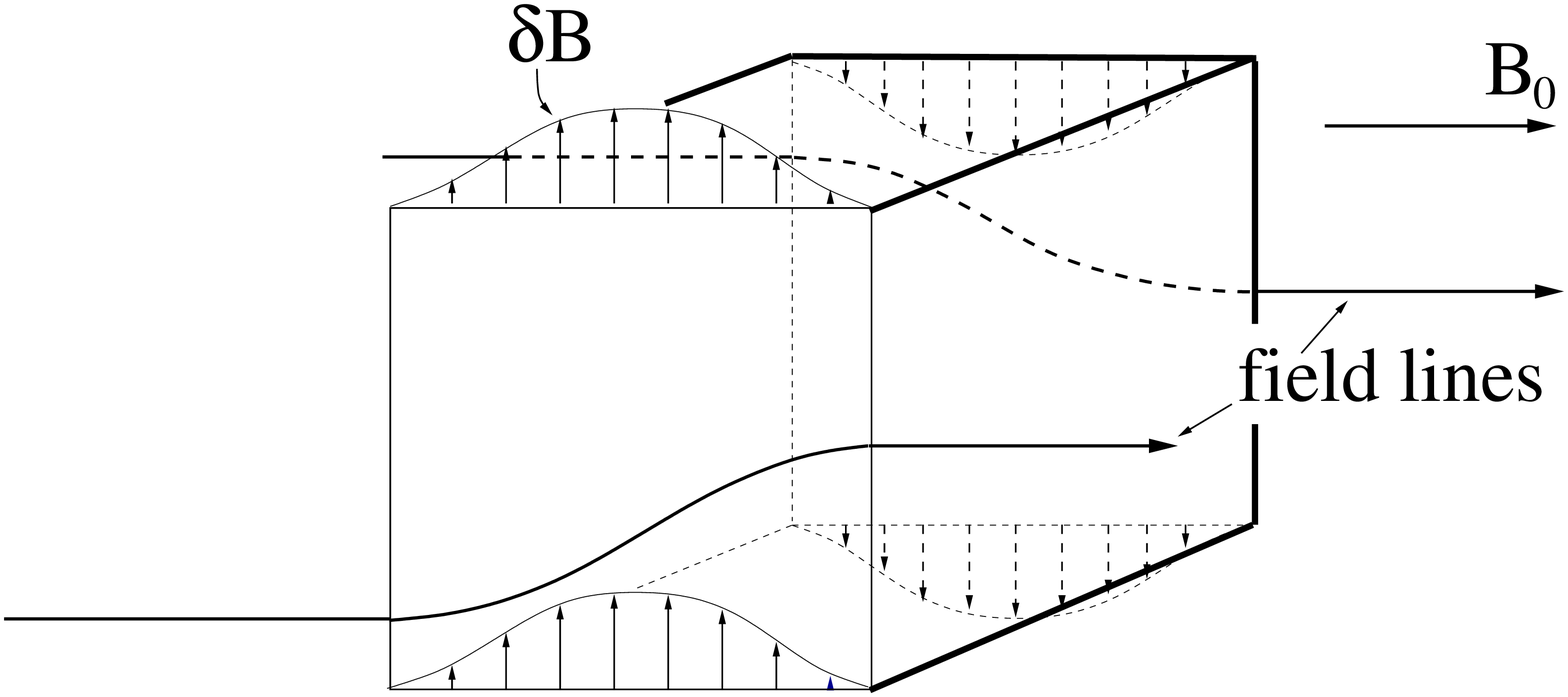}
\caption{An example of a cubical portion of a 3D wave packet.}
\label{fig:wavepacket2} 
\end{figure}

There are two types of waves in incompressible MHD, the Alfv\'en wave
and the slow wave, with polarizations illustrated in
figure~\ref{fig:polarizations}. Both waves propagate along
(or anti-parallel to) ${\bf B}_0$ at speed~$v_{\rm A}$. A 1D ${\bf
a}^-$-Alfv\'en-wave-packet (with ${\bf v} = - \delta {\bf b}$) is illustrated
in figure~\ref{fig:wavepacket1}.  A cubical portion of a 3D
wave packet is illustrated in figure~\ref{fig:wavepacket2}.


In general, a wave packet is associated with a displacement~$\triangle
{\bf r}$ of a field line, as in figures~\ref{fig:wavepacket1}
and~\ref{fig:wavepacket2}. For a small-amplitude wave packet that
perturbs the field lines only slightly,
\begin{equation}
\triangle {\bf r} = \frac{1}{B_0} \int_{-\infty}^\infty {\bf B}_\perp(z) \,dz
= \frac{\tilde{\bf B}_\perp(k_z=0)}{B_0} ,
\label{eq:dx1} 
\end{equation} 
where ${\bf B}_\perp$ is the component of~$\delta {\bf B}$ in the
$xy$-plane, and $\tilde{\bf B}_\perp$ is the Fourier transform
of~${\bf B}_\perp$ in~$z$.  Thus, the field-line displacement is given
by the $k_z =0$ component of a wave packet.~\cite{ng96}  If an
Alfv\'en wave packet's length along the field is $\lambda_\parallel$,
its width across the field is~$\lambda_\perp$, and its amplitude is
$B_{\lambda_\perp}$, then equation~(\ref{eq:dx1}) implies that
\begin{equation}
\triangle r \sim \frac{\lambda_\parallel B_{\lambda_\perp}}{B_0}
\sim \frac{\lambda_\parallel a_{\lambda_\perp}}{v_{\rm A}},
\label{eq:dx3} 
\end{equation} 
where~$a_{\lambda_\perp}$ is the wave-packet amplitude in Elsasser
variables.

\section{Wave-wave interactions in incompressible MHD}
\label{sec:wavewave} 

\begin{figure}[h]
\vspace{4.1in}
\includegraphics{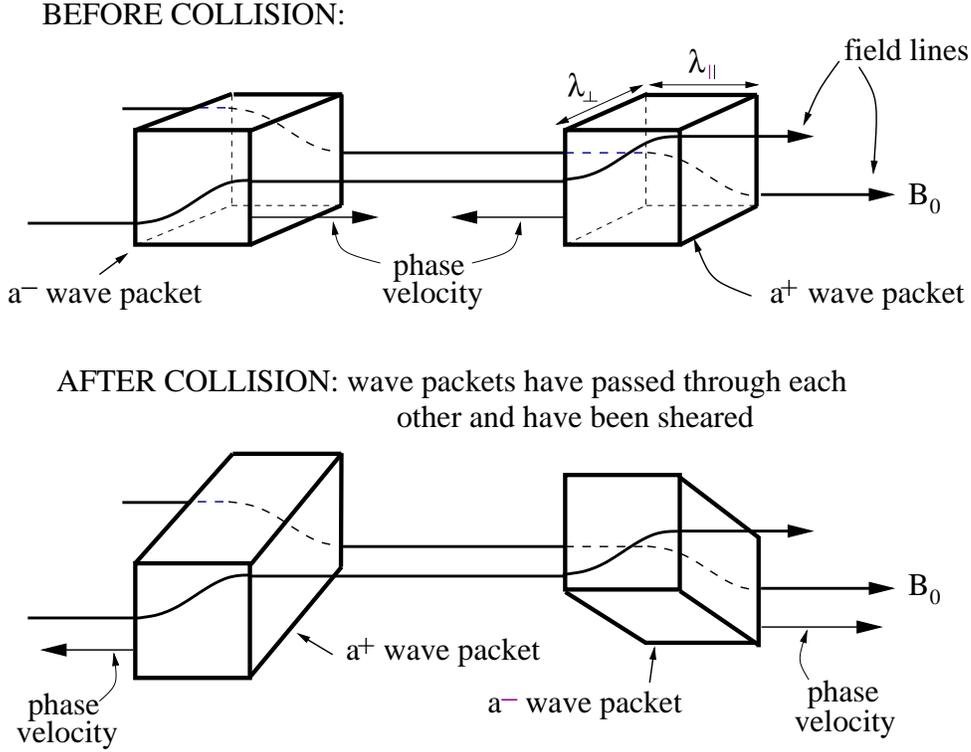}
\caption{During a wave-packet collision, the ${\bf a}^-$ (${\bf a}^+$) packet
follows the field lines of the 
${\bf a}^+$ (${\bf a}^-$) packet.}
\label{fig:collision1}
\end{figure}

For the nonlinear ${\bf a}^\pm \cdot \nabla {\bf a^{\mp}}$ terms in
equations~(\ref{eq:ap}) and (\ref{eq:am}) to be nonzero, both ${\bf
a}^+$ and ${\bf a}^-$ fluctuations must be present at the same point
in space. Wave-wave interactions can thus be thought of as collisions
between oppositely directed wave packets~\cite{kra65}.  When ${\bf
a}^+$ and ${\bf a}^-$ wave packets collide, to lowest order in wave
amplitude each wave packet follows the field lines associated with the
other wave packet~\cite{sri94,mar01}.  For example, before a collision
the value of~${\bf a} ^-$ remains constant at a point that moves along
${\bf B}_0$ at speed~$v_{\rm A}$.  During the collision, the value
of~${\bf a}^-$ remains approximately constant at a point that moves at
speed~$\sim v_{\rm A}$ along the field lines formed from the sum
of~${\bf B}_0$ and the field fluctuations of the~${\bf a}^+$ wave
packet, with a pressure-induced correction to ensure~$\nabla \cdot
{\bf a}^- =0$~\cite{mar01}.  This is illustrated in
figure~\ref{fig:collision1} for cubical portions of two colliding wave
packets. In both cubes, $\delta {\bf B}$ is upwards in the near-face
of the cube, and downward in the rear-face of the cube, as in
figure~\ref{fig:wavepacket2}.

The different displacements of different parts of a
wave packet during a collision
shear the wave packet in the plane perpendicular to~${\bf B}_0$.
For colliding wave packets of length~$\lambda_\parallel$ along ${\bf B}_0$
and width~$\lambda_\perp$ across ${\bf B}_0$,
the change in the value of~${\bf a}^\pm$ induced by the collision
is roughly the duration of a collision, $\lambda_\parallel/v_{\rm A}$,
times the magnitude of the nonlinear ${\bf a}^\mp\cdot \nabla {\bf a}^\pm$
term. If $|{\bf a}^+| \sim |{\bf a}^-| \sim a_{\lambda_\perp}$, and if
$\lambda_\perp \lesssim \lambda_\parallel$, then 
the fractional change~$\chi$ in~${\bf a}^\pm$ induced by the collision 
is given by~\cite{gol95}
\begin{equation}
\chi \sim \frac{a_{\lambda_\perp} \lambda_\parallel}{v_{\rm A} \lambda_\perp},
\label{eq:chi} 
\end{equation} 
or, from equation~(\ref{eq:dx3}), 
\begin{equation}
\chi \sim \frac{\triangle r}{\lambda_\perp}.
\label{eq:chi2} 
\end{equation} 

When $\chi \ll 1$, the right-hand side of, for example, the ${\bf
a}^-$ wave packet in figure~\ref{fig:collision1} is displaced in
almost the same way as the left-hand side, since both sides encounter
essentially the same~${\bf a}^+$ wave packet, since the~${\bf a}^+$
packet is only slightly altered during the collision.  Changes to the
profile of a wave packet along the magnetic field are thus weaker than
changes in the profile of a wave packet in the plane $\perp$~to~${\bf
B}_0$~\cite{she83,ng97,gol97}. As a result, the cascade of energy to
small~$\lambda_\parallel$ is much less efficient than the cascade of
energy to small~$\lambda_\perp$ in weak incompressible MHD turbulence,
as discussed in the following section.

\section{Weak incompressible MHD turbulence}
\label{sec:weak}

In weak incompressible MHD turbulence, $\chi \ll 1$ and nonlinear
wave-wave interactions are a small perturbation to a wave's linear
oscillatory behavior. In this limit, the wave kinetic equation for the
power spectra can be derived using weak turbulence theory~\cite{gal00}.
When only Alfv\'en waves are excited, the wave
kinetic equation is of the form~\cite{gal00}
\[
\frac{\partial E^\pm ({\bf k})}{\partial t}
= \int d^3 k_1 d^3 k_2 [N_{{\bf k}, {\bf k}_1, {\bf k}_2} E^\pm ({\bf k})-  P_{{\bf k}, {\bf k}_1, {\bf k}_2}
E^\pm ({\bf k}_1)] E^\mp({\bf k}_2) 
\]
\begin{equation} 
\times \delta(\omega_{{\bf k}}^\pm - \omega_{{\bf k}_1}^\pm -\omega_{{\bf k}_2}^\mp) 
\delta({\bf k} - {\bf k}_1 - {\bf k}_2),
\label{eq:wk} 
\end{equation} 
where ${\bf k}$ is the Fourier-space wave number,
\begin{equation}
E^\pm ({\bf k}) \delta ( {\bf k} + {\bf k}_1) = 
\langle \tilde{\bf a}^\pm ({\bf k}) \cdot \tilde{\bf a}^\pm ({\bf k}_1) \rangle ,
\label{eq:defE} 
\end{equation} 
$\tilde{\bf a}^\pm$ is the Fourier transform of~${\bf a}^\pm$,
$\langle \dots \rangle$ indicates an ensemble average,
the kernels $N_{{\bf k}, {\bf k}_1, {\bf k}_2}$ and $P_{{\bf k}, {\bf
k}_1, {\bf k}_2}$ are independent of~$E^\pm$ and given
by~\cite{gal00}, and
\begin{equation}
\omega_{{\bf k}}^\pm = \mp k_z v_{\rm A} 
\label{eq:disp} 
\end{equation} 
is the frequency of small-amplitude ${\bf a}^\pm$-waves, from
equations~(\ref{eq:ap}) and (\ref{eq:am}).  
Equation~(\ref{eq:wk}) contains only the leading-order terms in the weak-turbulence expansion,
called the three-wave-interaction terms. 
The delta functions in the
integrand give the three-wave resonance conditions,
\begin{equation} 
{\bf k} = {\bf k}_1 + {\bf k}_2,
\label{eq:1} 
\end{equation} 
and 
\begin{equation}
\omega_{{\bf k} }^\pm = \omega_{{\bf k}_1}^\pm + \omega_{{\bf k}_2}^\mp.
\label{eq:2} 
\end{equation} 
The integrand in equation~(\ref{eq:wk}) contains the product
$E^\pm E^\mp$ and no terms of the form $E^\pm E^\pm$, consistent
with the fact that only oppositely directed wave packets interact.
Upon dividing equation~(\ref{eq:2}) by $v_{\rm A}$, one finds that
\begin{equation} 
-k_z = -k_{1z} + k_{2z}
\label{eq:3} 
\end{equation} 
Combining equation~(\ref{eq:3}) with
the $z$-component of equation~(\ref{eq:1}) yields
\begin{equation} 
k_{2z} = 0, \mbox{ \hspace{0.3cm} and \hspace{0.3cm} }
 k_{z} = k_{1z}.
\label{eq:k1k3} 
\end{equation} 
If turbulence is not excited initially at some value of~$k_z$,
three-wave interactions will not excite turbulence at that value
of~$k_z$. Indeed, three-wave interactions do not transfer energy to
larger values of~$k_z$, only to larger values of $k_\perp =
\sqrt{k_x^2 + k_y^2}$~\cite{she83,gal00}. Also, the rate of energy transfer
for $E^\pm$  is in essence proportional to the $k_z=0$ part of~$E^\mp$.
These conclusions remain valid when slow waves are also excited.~\cite{gal00}

What physical meaning can be attributed to the $k_z=0$ part of the
power spectrum? It is simpler to answer this question by working with
a 1D spectrum ($M$, defined below) obtained by integrating~$E^+$
over~$k_x$ and $k_y$.  Assuming homogeneous turbulence, let
\begin{equation}
\langle {\bf B}^+(x,y,0) \cdot {\bf B}^+(x,y,z)\rangle = g(z) ,
\label{eq:defg} 
\end{equation} 
where ${\bf B}^+$ is the fluctuating magnetic  field associated with all 
~${\bf a}^+$ wave packets (${\bf B}^+  = {\bf a}^+\sqrt{\pi \rho}$),
and let 
\begin{equation}
M(k_z) = \int_{-\infty}^{\infty}  e^{-ik_z z} g(z) dz.
\label{eq:defM} 
\end{equation} 
Equations~(\ref{eq:defE}) and (\ref{eq:defM}) imply that $M(k_z) =
[\rho/32\pi^4]\int dk_x dk_y E^+ ({\bf k})$.
Assuming that all wave packets have coherence
length~$\lambda_\parallel$ along the magnetic field, coherence length
$\lambda_\perp$ across the field, and rms fluctuating field~$ B_{\lambda_\perp}$,
one finds that
\begin{equation}
M(0) = \left\langle \int_{-\infty}^{\infty} {\bf B}^+(x,y,0) \cdot {\bf
B}^+(x,y,z) dz \right\rangle \sim B_{\lambda_\perp}^2 \lambda_\parallel
.
\label{eq:M2} 
\end{equation} 
If the wave packets have the Alfv\'en-wave polarization, $\delta {\bf B} \perp {\bf B}_0$,
then  from equation~(\ref{eq:dx3}),
\begin{equation}
M(0)  \sim B_{\lambda_\perp} B_0 \triangle r,
\label{eq:M3} 
\end{equation} 
where $\triangle r \sim B_{\lambda_\perp} \lambda_\parallel/B_0$ is
the field-line displacement associated with a single wave packet.
Thus, for Alfv\'en waves, $M(0)$ [and similarly $E^\pm (k_z=0)$] is a
measure of the field-line displacements caused by wave packets.  The
dependence of the cascade rate on $E^\mp(k_z=0)$ indicates that energy
cascade arises from the turbulent wandering of field lines, as
described in section~\ref{sec:wavewave}.~\cite{ng96,ng97,gol97,lit01,lit03} As
illustrated by equation~(\ref{eq:M3}), the $k_z=0$ part of the
spectrum should not be equated with wave packets that have infinite
coherence length along the magnetic field~\cite{gal02,lit03}; structures with
finite~$\lambda_\parallel$ lead to nonzero~$\triangle r$, 
$M(0)$, and~$E(k_z=0)$.

The power spectrum of weak incompressible MHD turbulence can be
estimated as follows~\cite{ng97,gol97,gal00,bha01}. The fractional change~$\chi$
in a wave packet of width~$\lambda_\perp$ and
length~$\lambda_\parallel$ during one collision is~$\ll 1$.
Since the changes induced by successive collisions add randomly,
$\chi^{-2}$ collisions are required for an order-unity fractional
change. The time required for $\chi^{-2}$ collisions is $\chi^{-2}
\lambda_\parallel/v_{\rm A}$, or $ \sim v_{\rm A}
\lambda_\perp^2/(a_{\lambda_\perp}^2 \lambda_\parallel)$. This is
roughly the time $\tau_{\rm cascade}$ required for eddies of
width~$\lambda_\perp$ to pass their energy to eddies of width~$\sim
\lambda_\perp/2$. The cascade power~$\epsilon$ is roughly the energy
per unit mass in wave packets of width~$\lambda_\perp$,
i.e. $a_{\lambda_\perp}^2/2$, divided by~$\tau_{\rm cascade}$:
\begin{equation}
\epsilon \sim \frac{a_{\lambda_\perp}^4 \lambda_\parallel}{v_{\rm A} \lambda_\perp^2}.
\label{eq:eps} 
\end{equation} 
Assuming that local interactions in wave-number space dominate the
cascade, $\epsilon $ is independent of~$\lambda_\perp$  when
$\lambda_\perp \ll l$ and $\lambda_\perp \gg d$, where $l$ is the
outer scale (stirring scale) and $d$ is the dissipation scale. Since
$\lambda_\parallel$ is effectively constant,
\begin{equation}
a_{\lambda_\perp} \propto \lambda_\perp^{1/2}
\label{eq:aweak} 
\end{equation}
\cite{ng97,gol97,gal00}.
If the one-dimensional power spectrum $E_{1D}(k_\perp)$ is defined so that the
energy of modes with~$k_\perp$ in the interval~$(k_1,2k_1)$,
i.e. $a_{k_1^{-1}}^2$, 
is given by $\int_{k_1}^{2k_1} E_{1D}(k_\perp) dk_\perp \sim k_1 E_{1D}(k_1)$, 
then 
\begin{equation}
E_{1D}(k_\perp) \propto k_\perp^{-2}
\label{eq:eweak} 
\end{equation} 
\cite{ng97,gol97,gal00}.

Inserting equation~(\ref{eq:aweak}) into equation~(\ref{eq:chi}) yields~\cite{gol97}
\begin{equation}
\chi \propto \lambda_\perp^{-1/2}.
\label{eq:chiweak} 
\end{equation} 
Thus, if the dissipation scale~$d$ is sufficiently small, then at
sufficiently small~$\lambda_\perp$ the value of $\chi$ increases to~1,
and the turbulence becomes strong.

\section{Strong incompressible MHD turbulence}
\label{sec:strong}

In strong MHD turbulence, for which~$\chi \gtrsim  1$, fluctuations are not
 waves, since they are significantly distorted by nonlinear interactions
during a single wave period. Nor can propagation of fluctuations along 
the magnetic field be ignored, since, for example, for~$\chi \sim 1$ propagation is in some
sense equally important as wave-wave interactions. The combination of
propagation and nonlinear interaction determines the length of eddies
in strong incompressible MHD  turbulence as follows.~\cite{gol95,mar01,lit01}

\begin{figure}[h]
\vspace{2.4in}
\includegraphics{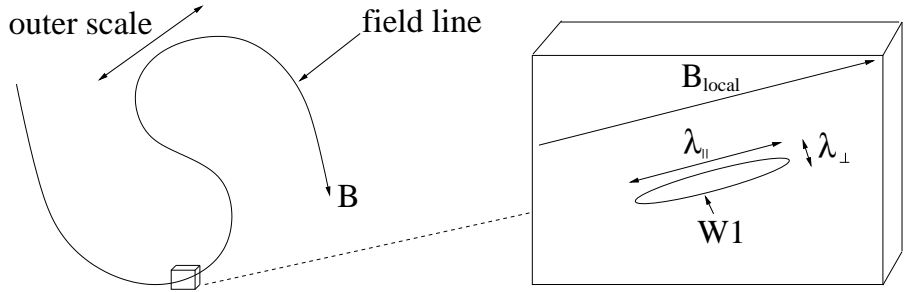}
\caption{In strong MHD turbulence, eddies are elongated along the local
magnetic field.}
\label{fig:la} 
\end{figure}

Consider a volume of width~$\lambda_\perp$ across the magnetic field
and length~$\lambda_\parallel $ along the field, with
$\lambda_\parallel \gg \lambda_\perp$.  The ${\bf a}^-$ fluctuations
within this volume will be loosely described as a ``wave packet'' with
the label~W1.  It is useful to consider the ``large-scale'' magnetic
field that is obtained by filtering out all magnetic fluctuations on
length scales comparable to or smaller than~W1. If one assumes that
the dominant nonlinear interactions are between structures of similar
size (local in $k$-space), one can neglect the shearing of~W1 by the
large-scale magnetic field. Within a box that is large compared to~W1
but small compared to the outer scale~$l$, one can thus approximate
the large-scale magnetic field as a uniform mean field denoted~${\bf
  B}_{\rm local}$, as in figure~\ref{fig:la}.  From
equation~(\ref{eq:am}), W1 propagates along ${\bf B}_{\rm local}$ at
speed $B_{\rm local}/\sqrt{4\pi \rho}$, while at the same time
undergoing nonlinear interactions. Let W2 be a ``wave packet''
of~${\bf a}^+$ fluctuations, also of width~$\lambda_\perp$ and length~$\lambda_\parallel$,
propagating antiparallel to ${\bf B}_{\rm
local}$ while undergoing nonlinear interactions. Let
$a_{\lambda_\perp}$ be the amplitude of ~${\bf a}^-$ within W1 and
also of ${\bf a}^+$ within W2.  Let $\lambda_{\parallel, \rm crit}$ be
the value of~$\lambda_\parallel$ such that~$\chi = 1$ for W1 and~W2:
$\lambda_{\parallel,\rm crit} = \lambda_\perp v_{\rm A}
/a_{\lambda_\perp}$.

\begin{figure}[h]
\vspace{2.3in}
\includegraphics{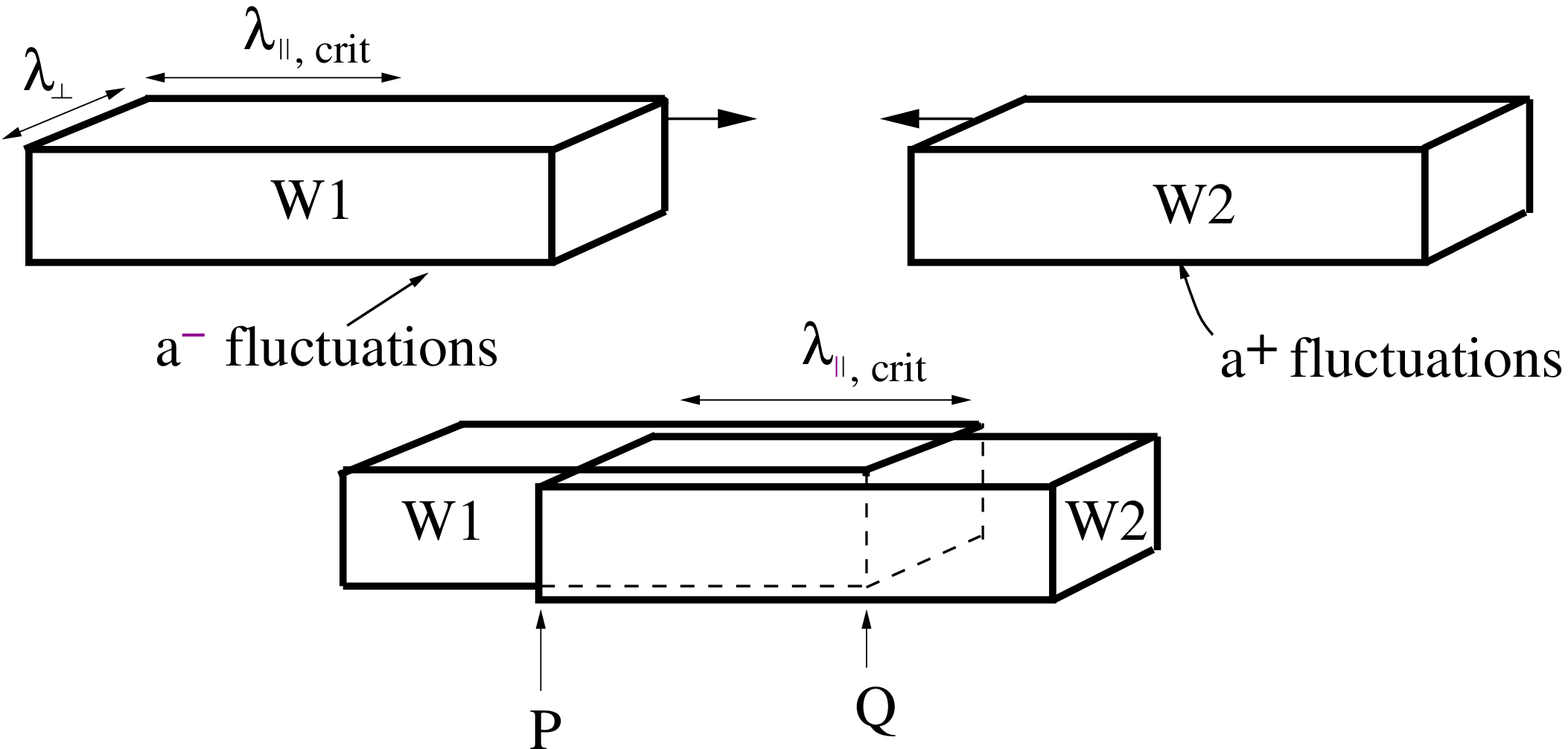}
\caption{}
\label{fig:coll2}
\end{figure}

Let us suppose that W1 and W2 collide, and that~$\lambda_\parallel\gg
\lambda_{\parallel, \rm crit}$ (i.e. $\chi \gg 1$), so that initially
the~${\bf a}^-$ and ${\bf a}^+$ fluctuations are coherent over a
distance~$\gg \lambda_{\parallel,\rm crit}$ along~${\bf B}_{\rm
  local}$. After W1 and W2 have propagated into each other a
distance~$\lambda_{\parallel, \rm crit}$, as depicted in the bottom
part of figure~\ref{fig:coll2}, the ~${\bf a}^-$ (${\bf a}^+$)
fluctuations at the leading edge of W1 (W2) above point Q (P) have
been altered by a factor of order unity by nonlinear interactions. As
time proceeds, the ${\bf a}^-$ fluctuations above points P and Q will
evolve in different ways, because they are interacting with
significantly different~${\bf a}^+$ fluctuations: the ${\bf a}^-$
fluctuations above point~Q interact with the pristine,
undistorted version of~W2; the ${\bf a}^-$ fluctuations above point~P
interact with ${\bf a}^+$ fluctuations in W2 that have already been
changed by the collision. As a result, the collision introduces structure
with parallel coherence length~$\lambda_{\parallel, \rm crit}$ into the two
wave packets.

According to these heuristic arguments, if $\chi$ is initially~$\gg 1$
(but finite),
the combination of nonlinear interaction and propagation along
field lines reduces~$\chi$ until~$\chi\sim 1$. On the other hand,
weak turbulence theory  shows shown that if~$\chi$ is initially~$\ll 1$,
nonlinear interactions act to increase~$\chi$ until~$\chi \sim 1$,
as discussed in section~\ref{sec:weak}.
Thus,  1~is in some sense the ``stable equilibrium value'' of~$\chi$
in incompressible MHD turbulence.~\cite{gol95} 

When $\chi \sim 1$, the turbulence is strong.
A theory for the spectrum and anisotropy of strong incompressible
MHD turbulence was developed by~\cite{gol95}.
The time $\tau_{\rm
cascade}$ for eddies of width~$\lambda_\perp$ and
length~$\lambda_\parallel$ to pass their energy to smaller eddies
is~$\sim \lambda_\parallel/v_{\rm A} \sim
\lambda_\perp/a_{\lambda_\perp}$. Assuming local interactions in
$k$-space, the cascade power~$\epsilon$ is given by
$a_{\lambda_\perp}^2/\tau_{\rm cascade} \sim
a_{\lambda_\perp}^3/\lambda_\perp$. Assuming that~$\epsilon$ is
independent of $\lambda_\perp$, 
\begin{equation}
a_{\lambda_\perp} \propto \lambda_\perp^{1/3},
\label{eq:as}
\end{equation}
which corresponds to
\begin{equation}
E_{1D}(k_\perp) \propto k_\perp^{-5/3}.
\label{eq:Es}
\end{equation}
The condition $\chi \sim 1$ combined with equation~(\ref{eq:as})
yields~\cite{gol95}
\begin{equation}
\lambda_\parallel \propto \lambda_\perp^{2/3}.
\label{eq:lp1}
\end{equation}
If the turbulence is stirred at scale~$l$, and if $a_l \simeq v_{\rm A}$,
then~\cite{gol95}
\begin{equation}
\lambda_\parallel = l^{1/3} \lambda_\perp^{2/3}.
\label{eq:lp2}
\end{equation}

Equations~(\ref{eq:as}) through~(\ref{eq:lp2}) apply
to both Alfv\'en-mode eddies and slow-mode eddies~\cite{gol95}.
However, Alfv\'en-mode fluctuations dominate the shearing of
both Alfv\'en-mode eddies and slow-mode eddies~\cite{gol95}.
The reason is that the nonlinear ${\bf a}^\pm \cdot \nabla
{\bf a}^\mp$ terms are much larger for Alfv\'en modes than
for slow modes, since fluctuations vary most rapidly across
the magnetic field, and ${\bf a}^\pm$ is perpendicular to
the local magnetic field for Alfv\'en modes, but nearly
along the magnetic field for slow modes when $\lambda_\perp\ll
\lambda_\parallel$.

Equations~(\ref{eq:as}) through~(\ref{eq:lp2}) are consistent with the
numerical results of~\cite{cho00,cho02}.  The numerical simulations
of~\cite{mul00} agree with equation~(\ref{eq:Es}), but do not address
local anisotropy. On the other hand, the numerical simulations
of~\cite{mar01} are consistent with equation~(\ref{eq:lp2}), but find
$E_{1D}(k_\perp) \propto k_\perp^{-3/2}$.  Thus, although
equations~(\ref{eq:as}) through~(\ref{eq:lp2}) are supported by
physical arguments and some direct numerical simulations, areas of
uncertainty remain.

\section{Decay and dynamic alignment}
\label{sec:dec} 

If only one type of fluctuation, either ${\bf a}^+$ or ${\bf a}^-$, is
present, nonlinear interactions vanish. Such a state is called
``maximally aligned,'' since either ${\bf v} = {\bf b} $ or ${\bf v} =
- {\bf b}$.  Decaying ``turbulence'' that is maximally aligned decays
only on a viscous or resistive time scale, and can thus be long-lived.

If decaying turbulence initially contains both ${\bf a}^+$ and ${\bf a}^-$
fluctuations, but an excess of one over the other, then it decays
to a maximally aligned state, in which only the initially predominating
fluctuation type remains.~\cite{dob80,gra83,tin86,gal00,mar01,cho02,lit03}
The reason for this is the following. Whether the turbulence is weak or strong,
isotropic or anisotropic, the energy-cascade time of ${\bf a}^\pm$ fluctuations
at a perpendicular scale~$\lambda_\perp$ is proportional to some power
of the energy in ${\bf a}^\mp$ fluctuations at scale~$\lambda_\perp$. If turbulence
is excited with an excess of~${\bf a}^+$ waves, the decay time for the
${\bf a}^-$ waves is short compared to the decay time of the ${\bf a}^+$
waves. The ${\bf a}^-$ waves thus decay faster than the ${\bf a}^+$ 
fluctuations,  with a discrepancy in decay rates that increases in time.
It has been suggested by~\cite{cho03} that this mechanism also operates
in compressible MHD turbulence, in which case it 
provides a potential mechanism for long-lived turbulence
in astrophysical environments.

\end{article}

\begin{thebibliography}{99} 

\bibitem{iro63} Iroshnikov, P. 1963, Astron. Zh., 40, 742

\bibitem{kra65} Kraichnan, R. H. 1965, Phys. Fluids, 8, 1385

\bibitem{pou76} Pouquet, A., Frisch, U., \& L\'eorat, J. 1976, J. Fluid Mech., 77, 321

\bibitem{dob80} Dobrowolny, M., Mangeney, A., \& Veltri, P. 1980,
Phys. Rev. Letters, 45, 144

\bibitem{mon81}Montgomery, D., \& Turner, L. 1981, Phys. Fluids, 24, 825

\bibitem{gra83} Grappin, R.,  Pouquet, A., L\'eorat, J. 1983,
Astron. Astrophys., 126, 51

\bibitem{hig84} Higdon, J. C. 1984, ApJ, 285, 109

\bibitem{she83} Shebalin, J. V., Matthaeus, W., \& Montgomery, D. 1983, J. Plasma Phys., 29, 525

\bibitem{tin86} Ting, A., Matthaeus, W., \& Montgomery, D. 1986, Phys. Fluids, 29, 3261

\bibitem{zan93} Zank, G. P., \& Matthaeus, W. H. 1993, Phys. Fluids, A5, 257

\bibitem{sri94} Sridhar, S., \& Goldreich, P. 1994, ApJ, 432 612

\bibitem{oug94} Oughton, S., Priest, E., \& Matthaeus, W. 1994, J. Fluid Mech., 280, 95

\bibitem{gol95} Goldreich, P., \& Sridhar, S. 1995, ApJ, 438, 763

\bibitem{pol95} Politano, H., Pouquet, A., \& Sulem, P.-L. 1995, Phys. Plasmas, 2, 2931

\bibitem{mon95} Montgomery, D., \& Matthaeus, W. 1995, ApJ, 447, 706

\bibitem{ng96} Ng, C. S., \& Bhattacharjee, A. 1996, ApJ, 465, 845

\bibitem{ng97} Ng, C. S., \& Bhattacharjee, A. 1997, Phys. Plasmas, 4, 605

\bibitem{gol97} Goldreich, P., \& Sridhar, S. 1997, ApJ, 485, 680

\bibitem{gho97} Ghosh, S., \& Goldstein, M. 1997, J. Plasma Phys., 57, 129

\bibitem{sto98} Stone, J., Ostriker, E., \& Gammie, C. 1998, ApJ, 508, L99 

\bibitem{kin98} Kinney, R. M., \& McWilliams, J. C. 1998, Phys. Rev. E, 57, 7111

\bibitem{bal99} Balsara, D., \& Pouquet, A. 1999, Phys. Plasmas, 6, 89

\bibitem{mac99} Mac Low, M. M. 1999, ApJ, 524, 169

\bibitem{vaz00} Vazquez-Semadeni, E., Ostriker, E., Passot, T., Gammie, C., \& Stone, J. 2000,
Protostars and Planets IV, eds Mannings, V., Boss, A.P., and Russell, S.
(Tucson: University of Arizona Press), p. 3

\bibitem{gal00} Galtier, S., Nazarenko, S. V., Newell, A. C., \& Pouquet,
A. 2000, J. Plasma Phys., 63, 447

\bibitem{mul00} M\"{u}ller, W., \&  Biskamp, D., 2000, Phys. Rev. Lett., 84, 475

\bibitem{cho00} Cho.,  J., \& Vishniac, E. 2000, ApJ, 539, 273

\bibitem{bha01} Bhattacharjee, A., \& Ng, N. 2001, ApJ, 548, 318

\bibitem{mar01} Maron, J., \& Goldreich, P. 2001, ApJ, 554, 1175

\bibitem{lit01} Lithwick, Y., \& Goldreich, P. 2001, ApJ, 562, 279

\bibitem{mil01} Milano, L., Matthaeus, W., Dmitruk, P., Montgomery, D. C. 2001,
Phys. Plasmas,  8, 2673

\bibitem{gal02} Galtier, S., Nazarenko, S., Newell, A., \& Pouquet, A. 2002, ApJ, 564, L49

\bibitem{oss02} Ossenkopf, V., \& Mac Low, M. 2002, A\& A, 390, 3070

\bibitem{sch02} Schekochihin, A., Maron, J., \& Cowley, S. 2002, ApJ, 576, 2002

\bibitem{bol02} Boldyrev, S., Nordlund, A, \& Padoan, P. 2002, ApJ, 573, 678

\bibitem{cho02} Cho, J., Lazarian, A., \& Vishniac, E. 2002, ApJ, 564, 291

\bibitem{cho03} Cho, J., \& Lazarian, A. 2003, astro-ph/0301062

\bibitem{win03} Winters, W., Balbus, S., Hawley, J. 2003, MNRAS, 340, 519

\bibitem{ves03} Vestuto, J., Ostriker, E., Stone, J. 2003, ApJ, 590, 858

\bibitem{pas03} Passot, T., \& Vázquez-Semadeni, E. 2003, A\&A, 398, 845

\bibitem{lit03} Lithwick, Y., \& Goldreich, P. 2003, ApJ, 582, 1220

\end{thebibliography}
\end{document}